\documentclass[twocolumn]{aastex62}
\usepackage{dcolumn} 
\usepackage{verbatim}
\usepackage{mathtools}
\usepackage{latexsym}
\usepackage{natbib}
\usepackage{amsmath}
\usepackage{amssymb}   
\usepackage{booktabs}
\usepackage{bm}
\usepackage{graphicx}
\usepackage{url}
\usepackage{color}

\newcommand{\be}{\begin{equation}}
\newcommand{\ee}{\end{equation}}
\newcommand{\ba}{\begin{eqnarray}}
\newcommand{\ea}{\end{eqnarray}}

\newcommand{\bi}{\begin{itemize}}
\newcommand{\ei}{\end{itemize}}
\newcommand{\mpch}{h^{-1} {\rm Mpc}}

\newcommand{\hmpc}{h {\rm Mpc}^{-1}}

\newcommand{\subp}{\mathrm{p}}
\newcommand{\cov}{\mathbf{C}}
\newcommand{\datap}{\mathbf{p}}

\begin{document}
\title{The Weak Lensing Peak Statistics in the Mocks by the inverse-Gaussianization Method}

\author[0000-0002-2183-9863]{Zhao Chen}
\affiliation{Department of Astronomy, Shanghai Jiao Tong University,
800 Dongchuan Road, Shanghai 200240, China}

\author[0000-0002-9359-7170]{Yu Yu}
\affiliation{Department of Astronomy, Shanghai Jiao Tong University,
800 Dongchuan Road, Shanghai 200240, China}
\email{yuyu22@sjtu.edu.cn}

\author{Xiangkun Liu}
\affiliation{South-Western Institute for Astronomy Research, Yunnan University Chenggong District, Kunming 650500, Yunnan Province, China}

\author{Zuhui Fan}
\affiliation{South-Western Institute for Astronomy Research, Yunnan University Chenggong District, Kunming 650500, Yunnan Province, China}
\affiliation{Department of Astronomy, School of Physics, Peking University, Beijing 100871, China}


\begin{abstract}
We apply the inverse-Gaussianization method proposed in \citealt{2016PhRvD..94h3520Y} to fast produce weak lensing convergence maps and investigate the peak statistics, including the peak height counts and peak steepness counts, in these mocks.
We find that the distribution of peak height and steepness is in good agreement with the simulation.
The difference is $\lesssim 20\%$ for these peak statistics in the maps at source redshift $z_s=1$.
Besides, the loss of off-diagonal elements in peak covariance motivates us to consider the super sample variance in weak lensing peak statistics.
We propose correction methods to effectively recover the (anti-)correlation among different bins by adding scatters in the mean value of these mocks. Finally, as an example of the application, we adopt the improved inverse-Gaussianization method with the above improvement to fast generate 40,000 mocks to calculate precision matrices between the power spectrum and peak statistics.
\end{abstract}



\section{Introduction}
\label{sec1}

The gravity induced deflection of the light from distant background galaxies provides an effective cosmological probe which is referred to weak gravitational lensing.
The lensing effect is induced by the total matter along the line-of-sight, no matter it is dark or luminous.
This makes weak gravitational lensing a promising tool to probe the matter distribution and even the geometry of the universe (e.g. \citealt{mellier1999probing,2001PhR...340..291B, 2008PhR...462...67M, Kilbinger2015Cosmology}).
Various weak lensing observations have improved our knowledge on the universe.
(e.g. CFHTLS: \citealt{2006A&A...452...51S, 2008A&A...479....9F};
CFHTLenS: \citealt{2013MNRAS.433.3373V, 2014arXiv1406.1379H};
KiDS-450: \citealt{2017MNRAS.471.4412K, 2017MNRAS.465.1454H};
DES Y1 results: \citealt{2018PhRvD..98d3528T, 2018PhRvD..98d3526A}).
Ongoing and upcoming projects, such as DES (\citealt{Abbott2018The}), KiDS (\citealt{De2015The}), HSC (\citealt{2018PASJ...70S...8A}), LSST (\citealt{2009arXiv0912.0201L}), Euclid (\citealt{Laureijs2011Euclid}), will obtain a huge amount of weak lensing observational data with unprecedented precision.
To match this statistical power, the control of various weak lensing systematics in the analysis is of great importance.

The weak lensing field is a non-Gaussian field due to the nonlinear evolution of the late Universe (e.g. \citealt{Scoccimarro1999Power, Lee2008Information, Semboloni2010Cosmic, Takada2010The, Joachimi2011Cosmological}).
The second order statistics, i.e. the correlation function and the power spectrum, only capture a part of the cosmological information.
Thanks to the improved statistical power, extracting the information in high order statistics helps to break the degeneracy in the cosmological parameters
(e.g. \citealt{2014MNRAS.441.2725F, 2015PhRvD..91j3511P, 2019arXiv191105568G}).
The weak lensing bispectrum and peak statistics are widely used in the analysis beyond the second order (e.g. \citealt{2009arXiv0912.0201L,  2010ApJ...719.1408F, 2011PhRvD..84d3529Y, 2016PhRvD..94d3533L, 2016PhRvL.117e1101L, 2016MNRAS.463.3653K, Shan2017KiDS, 2018MNRAS.478.5436G, 2018MNRAS.473.3190H, Martinet2018KiDS,2019JCAP...05..043C, 2019arXiv191004627M}).

In weak lensing peak analysis, a peak is defined as a local maximum whose value is greater than the eight neighbors in the simulated/observed convergence field.
Usually smoothing is adopted to suppress the noise in observation and simulation.
High peaks are mainly dominated by the lensing contribution from a single massive halo along the line-of-sight, while the low peaks are dominated by the contributions of multiple haloes close to each other in angular position (e.g. \citealt{2018ApJ...857..112Y,2018MNRAS.478.2987W}).
Separating the peaks into high and low type helps in peak modeling due to the different origins of these peaks.
Both of them contain cosmological information (e.g. \citealt{Hamana2010Searching, 2010MNRAS.402.1049D, 2010PhRvD..81d3519K, 2015A&A...576A..24L, 2016PhRvD..94d3533L}).
The number of peaks with different peak values is an important statistic called peak (height) counts which is sensitive to the non-Gaussianity of the lensing field.
\citealt{2011PhRvD..84d3529Y} found that the joint analysis of the power spectrum and peak height distribution is approximately twice sensitive in the cosmological parameter $\Omega_m$ and $\sigma_8$ compared to using power spectrum alone.
\citealt{2015PhRvD..91f3507L} used the CFHTLenS observations to constrain cosmology in a three-dimensional parameter space, $\Omega_m$, $\sigma_8$ and the dark energy equation-of-state $\omega$, and the constraint from peak height counts is comparable to those from the second order statistics alone. \citealt{Shan2017KiDS} indicated the peak height counts have the potential to break the degeneracy between $\Omega_m$ and $\sigma_8$ from the recent KiDS-450 observations (e.g. \citealt{2017MNRAS.465.1454H, Martinet2018KiDS}).
Counting the weak lensing peaks is one of the most promising statistics to extract information beyond Gaussian case.

Motivated by the recent deep learning study on weak lensing, another peak statistic, the peak steepness counting, was inferred. 
\citealt{Ribli2018Learning} found that the constraining power from the peak steepness counts is much greater  than the peak height counts in both noiseless and noisy case. 
Now the peak steepness is of great interest by the lensing community.

To obtain the precise cosmological parameter constraints, an accurate estimation of the covariance matrix is crucial.
The number of realizations to ensure $1\%$ error in the power spectrum covariance is estimated to be $\sim \mathcal{O}(10^4)$ (e.g. \citealt{2013PhRvD..88f3537D,2013MNRAS.432.1928T,2014MNRAS.439.2531P}).
Constructing such a large amount of realizations is a challenge especially for weak lensing cosmology, which requires both the large volume coverage and the small scale precision.
Fast simulations are widely adopted in galaxy clustering measurement (e.g. \citealt{2015MNRAS.446.2621C,2017JCAP...10..003A,2019MNRAS.482.1786L,2019MNRAS.485.2806B,2019MNRAS.482.4883C}).
Some of them are further developed for weak lensing (e.g. \citealt{2018MNRAS.473.3051I}).
Recently, \citealt{Peel2016Cosmological} studied the application of a fast model, CAMELUS, for peak height counts prediction in the Euclid-like survey. 
Machine Learning technique was also adopted to fast produce the weak lensing maps (e.g. \citealt{2019ComAC...6....1M}). 
Based on the fact that the weak lensing field can to effectively Gaussianized by a local transform (\citealt{2011PhRvD..84b3523Y}),
\citealt{2016PhRvD..94h3520Y} proposed the inverse-Gaussianization method to fast generate lensing mocks.
This method successfully produces the lensing power spectrum and a reasonable good power spectrum covariance.
The bispectrum amplitude is also recovered although the detailed dependence on the wavenumber configuration is missing.

In this work, we extend the analysis in \citealt{2016PhRvD..94h3520Y}.
We adopt the inverse-Gaussianization method to fast generate weak lensing convergence maps and quantify the statistics of the convergence peaks including peak height and peak steepness counts.
This paper is organized as follows. In Section \ref{sec2}, we review some basic knowledge about weak gravitational lensing and the inverse-Gaussianization method. 
Then, weak lensing convergence mocks are generated by the fast generating method and the peak counts results are presented in Section \ref{sec3}.
We briefly introduce an application of the mocks produced by the inverse-Gaussianization method in Section \ref{sec4}. 
We present the conclusion and discussion in Section \ref{sec5}.

Note that \citealt{2017MNRAS.465.1974S} adopted a similar local transform method to produce weak lensing maps and concluded that
the local-Gaussianized model cannot explain the simulated peak counts in the range of $\kappa/\sigma_\kappa<3$. The difference is about $10\%$ and increases towards low peaks.
We compare our results to \citealt{2017MNRAS.465.1974S} in Section \ref{sec3}.


\section{ The Inverse-Gaussianization Method }
\label{sec2}

\subsection{Weak Lensing Basics}

Weak lensing is one of the important cosmic probes at the late-time Universe. 
From the distortion of the background galaxy images, we can probe the lensing effect from the total matter along the line-of-sight.
The lensing effect is described by the Jacobi matrix
\be
\label{eqn:Jacobi}
\mathcal{A} = \left( \begin{array}{cc}{1-\kappa-\gamma_{1}} & {-\gamma_{2}} \\ {-\gamma_{2}} & {1-\kappa+\gamma_{1}}\end{array} \right)\ .
\ee
Here, $\kappa$ is the lensing convergence field which describes the amplification of the observed object.  While the complex shear, $\gamma=\gamma_{1}+\mathrm{i} \gamma_{2}$, gives the shape change of the image.  
They are not independent observables and can be converted to each other in the ideal case.
In this work, we focus on the lensing convergence, which is a weighted projection of the matter density along the line-of-sight,
\be
\kappa(\hat{n}, z_{s})=\int_{0}^{\chi_{s}} W(z, z_{s}) \delta_m(\hat{n}, z) d \tilde\chi.
\label{eqn:kappa}
\ee
Here, $\delta_m$ is the matter density contrast.  $\chi(z)$ is the comoving angular diameter distance to the lens redshift $z$.
We conventionally express $\chi$ in units of the Hubble radius, $\tilde\chi\equiv\chi/(c/H_0)$, in which $H_0$ is the Hubble constant today.
The lensing kernel for a source at redshift $z_s$ and a lens at redshift $z$ is given by
\be
W\left(z, z_{s}\right)=\frac{3}{2} \Omega_{m}(1+z) \tilde\chi(z) \left[ 1-\frac{\chi(z)}{\chi(z_s)}\right]\ .
\label{eqn:W}
\ee

In observation, a common way to gain the convergence field is to reconstruct it from the observed shear field (e.g. \citealt{2017ApJ...839...25S}). 
Because of the intrinsic shape noise, the reconstructed convergence is noisy, and usually we smooth the convergence field by using a Gaussian smoothing filter (e.g. \citealt{Hamana2010Searching}).
\be 
\mathcal{K}_{smoothed}(\theta)=\int \mathrm{d}^{2} \theta^{\prime} \kappa_{}\left(\theta-\theta^{\prime}\right) U\left(\theta^{\prime}\right),
\label{eqn:smooth}
\ee 
where $U(\theta)$ is the filter function:
\be
U(\theta)=\frac{1}{\pi \theta_{\mathrm{G}}^{2}} \exp \left(-\frac{\theta^{2}}{\theta_{\mathrm{G}}^{2}}\right).
\label{eqn:Utheta}
\ee
Here, we choose the smoothing scale $\theta_G = 1.2\text{arcmin}$, which is a suitable scale for the source at $z_s = 1.0$.

The two point correlation function and its Fourier pair power spectrum are the basic statistical tools in cosmology. The power spectrum is defined as the correlation in the Fourier space:
\be
\langle\kappa(\vec{\ell}_{1}) \kappa(\vec{\ell}_{2})\rangle=(2 \pi)^{2} \delta_{D}(\vec{\ell}_{1}+\vec{\ell}_{2}) C_{\kappa}(\ell_{1}).
\label{eqn:ps}
\ee
Here, Dirac function $\delta_D$ represents the homogeneity in statistics. The isotropy is implied by the fact that the power spectrum $C_{\kappa}(\vec{\ell})$ only depends on the mode $\ell=|\vec{\ell}|$.  By using the Limber approximation and Eq.\ref{eqn:kappa}, the convergence power spectrum is related to the matter power spectrum:
\be 
C_{\kappa}(\ell)=\int_{0}^{\chi_s} \mathrm{d} \chi \frac{W(\chi,\chi_s)}{\chi^2} P_{\delta}\left(k=\frac{\ell}{\chi} ; \chi\right)\ ,
\label{eqn:ps2}
\ee
and contains important cosmological information.

The weak lensing two point statistics have the most constrain power for the cosmological parameter combination $S_8=\sigma_8\Omega_m^{\alpha}$ with some power index $\alpha$ depending on the observation configuration and fiducial cosmology.
To break the degeneracy, high-order statistics are helpful.
The weak lensing skewness and kurtosis have been already measured and studied for more than ten years (e.g. \citealt{2003ApJ...598..818Z, 2004MNRAS.352..338J, 2009A&A...505..969P,2012MNRAS.423..983P,2019arXiv191105568G}). 
Analog to the power spectrum, the bispectrum $B_{\kappa}$ is defined by the three point correlation function of convergence: 
\begin{equation}
\begin{aligned}
\langle
\kappa(\vec{\ell_{1}})
\kappa(\vec{\ell_{2}})
\kappa(\vec{\ell_{3}})
\rangle 
=(2\pi)^{2}\delta_{\mathrm{D}}(\vec{\ell_{1}}+\vec{\ell_{2}}+\vec{\ell_{3}})
B_{\kappa}(\vec{\ell_{1}}, \vec{\ell_{2}},\vec{\ell_{3}}).
\end{aligned}
\end{equation}
The joint analysis of the two point statistics and three point statistics can tighter the constrain on $\Sigma_{8}=\sigma_{8}\left(\Omega_{\mathrm{m}} / 0.27\right)^{\alpha}$ \citep{2014MNRAS.441.2725F}.
Other than the high-order moments and the correlation hierarchy,
the weak lensing peak is considered as one of the most powerful statistics beyond the 2nd order (e.g. \citealt{2011PhRvD..84d3529Y, 2015A&A...583A..70L, 2016PhRvD..94d3533L}). 
In this paper, we focus on the peak statistics including the well studied peak height counts and the new developped peak steepness counts.

Hereafter, we only use the smoothed lensing convergence maps.
Thus, we still use the notation $\kappa$ to denote the smoothed lensing convergence field.

\subsection{Fast Mock Generation}
\label{sub:lf}
To capture more information from the non-Gaussian weak lensing field, several Gaussianization methods were proposed (e.g. \citealt{2011PhRvD..84b3523Y, Joachimi2011Cosmological, 2012ApJ...748...57S, 2012MNRAS.421..832Y}).
Based on the fact that the weak lensing field could be effectively Gaussianized (\citealt{2011PhRvD..84b3523Y}), the inverse-Gaussianization method was proposed to fast generate the lensing convergence maps (\citealt{2016PhRvD..94h3520Y}).
This method includes the following steps.

\bi

\item[ i ] \emph{Obtain the local transform function.}  Given a non-Gaussian lensing convergence map, we can find a monotonic local transform to map the $\kappa$ field into a new $y$ field.
We require that the $y$ field has a Gaussian one-point PDF by design and the local transform function is obtained by
\be 
y = \mathrm{erf}^{-1}(2\mathrm{CDF}(\kappa)-1).
\label{eqn:y}
\ee
Here, $\mathrm{CDF}(\kappa)$ is the cumulative distribution function of the lensing convergence field. The Gaussian error function $\mathrm{erf}(x)$ is given by the following expression:
\be
\mathrm{erf}(x)=\frac{2}{\sqrt{\pi}} \int_{0}^{x} e^{-\eta^{2}} d \eta .
\label{eqn:erf}
\ee 
The resulting $y$ field has the mean of zero and $\sigma_y = \sqrt{2}/2$.
For multiple realizations, we obtain the mean local transform first and use it as the Gaussian transform function to obtain the $y$ fields.

\item[ ii ] \emph{Obtain the realizations of y fields. }

From the Gaussianized $y$ maps, we can obtain the averaged power spectrum among the realizations.
New Gaussian $y$ realizations are generated according to this measured power spectrum.

\item[ iii ] \emph{Inverse transform.} 
Because we have obtained the local transform function ($y-\kappa$ relation) in Step i, it's straightforward to obtain the realizations of the weak lensing convergence maps by the inverse transform the Gaussian $y$ map. These independent realizations are what we need.

\ei

This method relies on the assumption that the weak lensing field could be effectively Gaussianized.
However, the residual non-Gaussianity is observed in the Gaussianized $y$ fields.
This residual non-Gaussianity definitely has some impact on some of the statistics.
We mainly focus on the peak statistics in this work.  

\subsection{ Simulation Setups}

The weak lensing convergence maps we use are from the simulation suite described in \citealt{2015MNRAS.450.2888L}.
The dark matter only simulations were run by Gadget-2  code \citep{2005MNRAS.364.1105S}, in the flat $\Lambda$CDM framework, with $\Omega_m=0.28$, $\Omega_\Lambda=0.72$, $\Omega_b=0.046$, $\sigma_8=0.82$, $n_s=0.96$ and $h=0.7$.
The simulation suite contains 8 independent realizations with a box size of $320\ \mpch$ and 4 independent realizations with a box size of $600\ \hmpc$.
For both cases, the particle number is $640^3$.
The lensing convergence maps at $z=1$ are constructed by the ray-tracing technique on the small box realizations.
Consider the different Cartesian directions, there are 24 sets in total and every set can be used to construct 4 maps.
Totally, 96 lensing convergence maps at $z=1$ are constructed with map area of $3.5\times 3.5\deg^2$, pixelized into $1024\times 1024$ grids.
We refer the readers to \citealt{2015MNRAS.450.2888L} for the details of the lensing simulation.


\section{ Results }
\label{sec3}
We quantify the performance of inverse-Gaussianization method against simulations through a series of tests.
Hereafter, we call the maps generated by the inverse-Gaussianization as lensing convergence \textit{mocks}.

\subsection{ Transform function }

\begin{figure}[ht!]
\centering
\includegraphics[width=8.5cm]{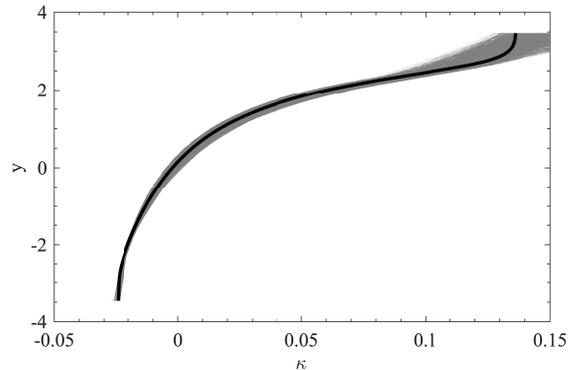}
\caption{ The local transforms ($\kappa$--$y$ relations) obtained from 96 N-body simulation realizations at $z_s = 1.0$ (grey lines). The black line in the middle is the averaged relation which is used to Gaussianize the lensing maps.
By design, the 96 $y$ maps all have the Gaussian PDF.
This operation is the first step of our local transform method, detailed in Sec.\ref{sec2}. }
\label{fig1}
\end{figure} 

The local transform functions ($\kappa$--$y$ relations) obtained from the 96 lensing convergence maps at $z=1$ are shown in grey lines in Fig. \ref{fig1}. 
The black line represents the average transform function.
We can see the divergence at the high end where the function is dominated by the rare extreme $\kappa$ values.  This phenomenon is caused by the sample variance. 

\subsection{ Power spectrum }

\begin{figure*}[ht!]
\centering
\includegraphics[width=8cm]{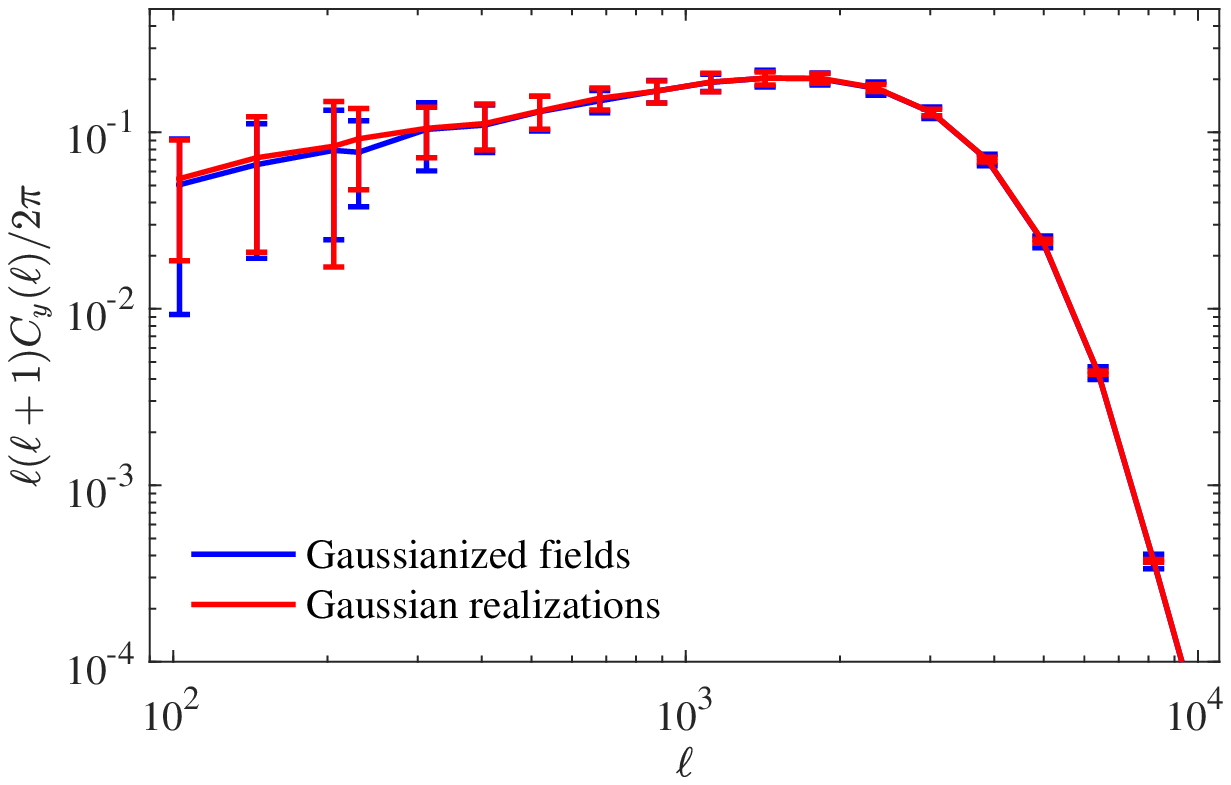}
\includegraphics[width=8cm]{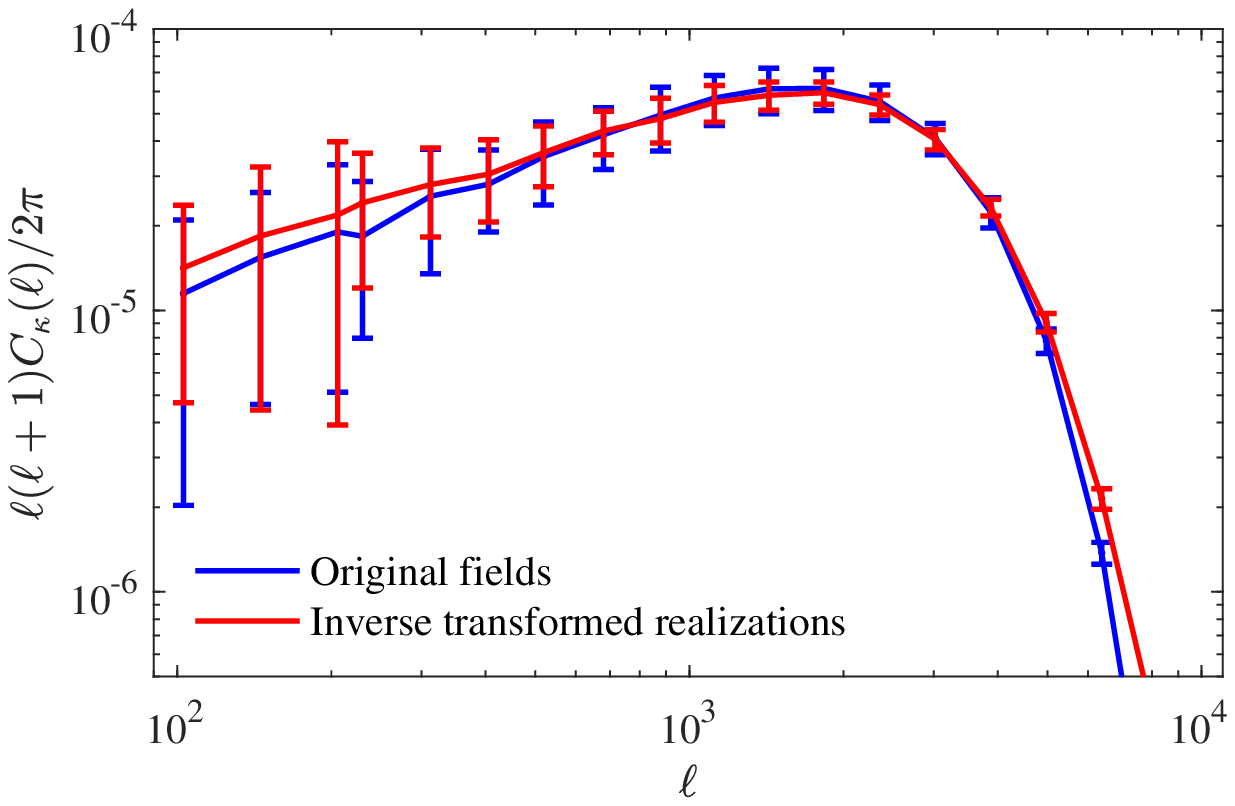}
\caption{ The blue line in the left plot represents the power spectrum $C_y(\ell)$ of Gaussianized $y$ fields which are produced at Step i in Sec.\ref{sub:lf}.
The error bars are the r.m.s. among realizations.
Convolving this measured $C_y(\ell)$ with white Gaussian Random Fields, we can obtain the Gaussian realizations whose power spectrum is shown as a red line in the left panel.
In the right panel we compare the power spectrum of simulated maps (blue line) with the $\kappa$ mocks produced by locally transforming the Gaussian Random Fields back (red line).
They are consistent with each other except at the very small scale.}
\label{fig:power}
\end{figure*}

First, we investigate the power spectrum of the lensing convergence mocks produced by the inverse-Gaussianization method.
The power spectrum from the simulated lensing maps is presented in blue line in the right panel of Fig. \ref{fig:power}.
The error bars are the r.m.s. among the 96 realizations.
Note that we applied Gaussian smoothing with $\theta_G = 1.2\text{arcmin}$ on these lensing maps.
The power at $\ell>3000$ is suppressed.
For the $y$ maps, we present the power spectrum in blue line in the left panel of Fig. \ref{fig:power}.
The power spectrum amplitude for these Gaussianzied fields is determined by $\sigma_y$ in the Gaussianization process.
However, the choice of $\sigma_y$ value has no impact on the mock power spectrum amplitude since the scaling is cancelled in the inverse-transform process.
We produce the Gaussian random fields with the measured power spectrum $C_y(\ell)$ for the same number of realizations.
The power is shown in red line in the left panel of Fig. \ref{fig:power}, which is consistent with the input power.
The power for the inverse-transformed fields is plotted in red line in the right panel.
We find the power spectrum is successfully reproduced by the inverse-Gaussianization method.
The slight difference at the largest scale comes from the cosmic variance.

In \citealt{2016PhRvD..94h3520Y}, the performance of the inverse-Gaussianization method is investigated on the projected density field over the thickness of $300\ \hmpc$ at several redshifts.
The maps used in this work is constructed from the ray-tracing method, which includes all the contribution from the nonlinear density field from $z=0$ to $z=1$ weighted by the lensing kernel $W(z,z_s=1)$.
The result here is consistent with the finding in \citealt{2016PhRvD..94h3520Y}.



\subsection{ Peak Height }

\begin{figure}[ht!]
\centering
\includegraphics[width=8.5cm]{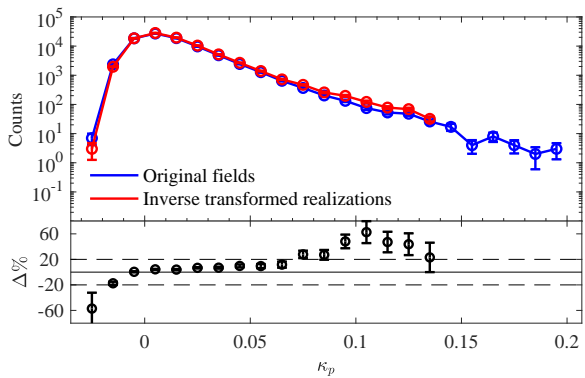}
\caption{ The peak height counts of the mocks and simulation realizations are shown in red and blue line in the top panel, respectively.  The error bars are generated by the bootstrap method.
The ratio of the two is shown in the bottom.
The peaks are reproduced well for $\kappa_\subp \in [-0.015,0.065]$.
The mocks from the inverse-Gaussianization method contain less peaks of extreme low value ($\kappa_\subp<-0.03$), more peaks with $\kappa_\subp\sim 0.1$, and miss the extremely high peaks ($\kappa_\subp>0.13$).
The missing of the extremely high peaks and the overproduction of the high peaks can be understood.
Adopting the averaged transform function $\kappa$--$y$ can not capture the map-to-map variance for these extremely high peaks and as a consequence they accumulate as the overproduction of high peaks.}
\label{fig:peak}
\end{figure}

Peak height counts of the weak lensing convergence map is considered as a promising tool to probe nonlinear structure evolution at late times, which can provide additional cosmological information beyond power spectrum/correlation function.  We are interested in whether the inverse-Gaussianization method could recover the peaks of the lensing field.

We compare the peak height counts $N(\kappa_\subp)$ with peak value $\kappa_\subp$ between the mocks and simulations in Fig. \ref{fig:peak}.
We divide the peaks into linear $\kappa_\subp$-bins covering $-0.03<\kappa_\subp<0.20$.
In the upper panel, the blue line represents the peak counts of mocks while the simulation result is drawn by the red line.
The error bars come from the non-parametric bootstrap method.
We plot the ratio of the mock peak counts to the simulated one in the bottom panel.
The error bars come from the fluctuations in the mocks.
We find that the peak height counts are produced reasonably good by the inverse-Gaussianization method.
For $-0.015<\kappa_\subp<0.065$ ($-1.23<\nu\equiv\kappa_\subp/\sigma_{\kappa}<5.37$), the difference is within $20\%$.
The peaks with value $\kappa_\subp >0.065$ are overproduced.

Another obvious feature is that the inverse-Gaussianization method cannot produce the extremely high peaks ($\kappa_\subp>0.13$, or $\nu>11$).
Two facts could induce this difference.
One is that the Gaussianization process is not perfect.
In \citealt{2016PhRvD..94h3520Y}, residual non-Gaussianity is observed in the power spectrum covariance for the Gaussianized fields.
The residual locates at small scales and implies that the peaks in the Gaussianized field are different from the one in a true Gaussian random field.
The inverse-Gaussianization method cannot capture these peaks, leading to the missing high end in the peak height counts.
The other fact is that we use the averaged transform function to perform the inverse-transform, partially neglecting the variance among the realizations.
This also fails to convert some high peaks in $y$ mocks into extremely high peaks in $\kappa$ mocks.
Those underestimated peaks are accumulated around $\kappa_\subp\approx 0.1$ and this partially causes the overproduction of the peaks at this range.


This result is slightly different from the result in \citealt{2017MNRAS.465.1974S}.
We list several differences in the process.
\citealt{2017MNRAS.465.1974S} adopts the convergence maps reconstructed from the ray-tracing shear fields, while our convergence maps are directly from the ray-tracing output.
\citealt{2017MNRAS.465.1974S} adopts the truncated Gaussian filter to account for an undetermined constant in the convergence construction,
while we use a simple Gaussian filter for theoretical motivation.
The power spectrum of the Gaussianized $y$ field in \citealt{2017MNRAS.465.1974S} is theoretically obtained by its relations with the local transform function and the power spectrum of the $\kappa$ field.
To numerically solve for $C_y(\ell)$, a continuous $C_\kappa(\ell)$ is required and obtained by modifying the theoretical convergence power spectrum from the halo model with two tuning factors accounting for the ray-tracing resolution effects.
In our method, we directly measure $y$ power spectrum from the Gaussianized fields and use this measured $C_y(\ell)$ to fast produce new $y$ maps.
In principle, all the numerical and resolution effects are automatically included in our process.
Our method tends to slightly overproduce the number of peak for $0<\nu<5$, while \citealt{2017MNRAS.465.1974S} produces less peaks in this range.
For the absolute deviation from the simulated one, our method has a slightly better performance in the range $-1<\nu<2$.
Both methods fail in the number of the extremely low peaks.


\begin{figure}[ht!]
\centering
\includegraphics[width=8.5cm]{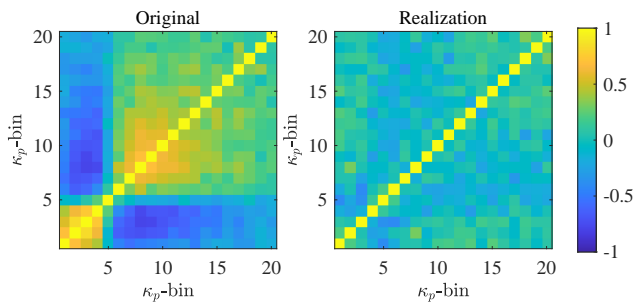}
\caption{ The normalized covariance matrix of peak height counts from simulation is shown in the left panel. We can find obvious structures which indicate the strong (anti-)correlation among different peak height bins. The right matrix is produced by inverse-Gaussianization method and is approximately diagonal. This result implies the absence of cosmological information caused by the super sample variance. }
\label{fig:covariance}
\end{figure}

\subsection{ Peak Height Covariance }
\label{subsec:Covariance}

From the 96 realizations for both the simulated and mock lensing convergence, we estimate the covariance of the peak height counts, $\cov_{ij}=\langle \Delta N(\kappa_{\subp,i})\Delta N(\kappa_{\subp,j})\rangle$.  Here, $i, j$ is the index for the peak value bin and $\Delta N(\kappa_{\subp,i})=N(\kappa_{\subp,i})-\bar{N}(\kappa_{\subp,i})$.
We choose 20 $\kappa_\subp$-bins linearly covering the range $-0.02<\kappa_\subp<0.1$.
The normalized peak height counts covariance 
\be
\hat{\cov}_{i j} = \frac{\cov_{i j}}{\sqrt{\cov_{i i} \cov_{j j}}}
\label{eqn:Cov}
\ee
is shown in Fig. \ref{fig:covariance}.
It presents the correlation among counts at different convergence peak values, which is important for constraining cosmological parameters using the peak height counts. The left panel is the peak covariance matrix of the simulated maps.
We observed strong off-diagonal features in this covariance.
The number of low peaks is strongly correlated with themselves, and the same happens for the high peaks.
The low peaks and high peaks have strong negative correlations.

However, the covariance of the mocks, which is shown in the right plot, is almost diagonal. 
This sharp contrast implies that there must be some important ingredient missing in the generation of the mocks.
Another clue is that this missing ingredient has almost no impact on the power spectrum.


\subsection{ Super Sample Variance }
\label{subsec:ssv}

\begin{figure}[ht!]
\centering
\includegraphics[width=8.5cm]{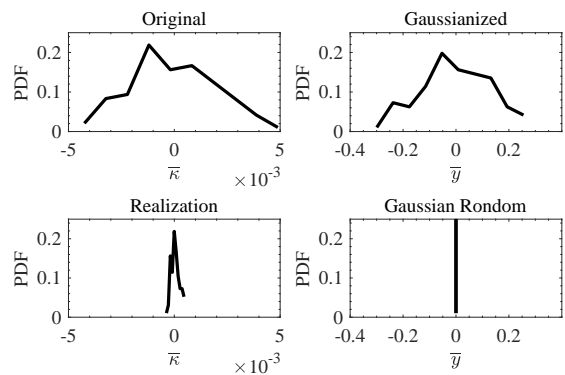}
\caption{ The distribution of the mean for convergence fields are shown in the left panels while the right panels represent the distribution of mean for $y$ fields. The original simulations and Gaussianized $y$ fields have broad scatter in the mean of fields.
However, the Gaussian Random Fields (newly generated $y$ fields, detailed in Section \ref{sub:lf}, Step ii) has exact zero mean.
It leads to the very narrow distribution of mean for these mocks, which differs from the simulation.}
\label{fig:mean}
\end{figure}

\begin{figure}[ht!]
\centering
\includegraphics[width=8.5cm]{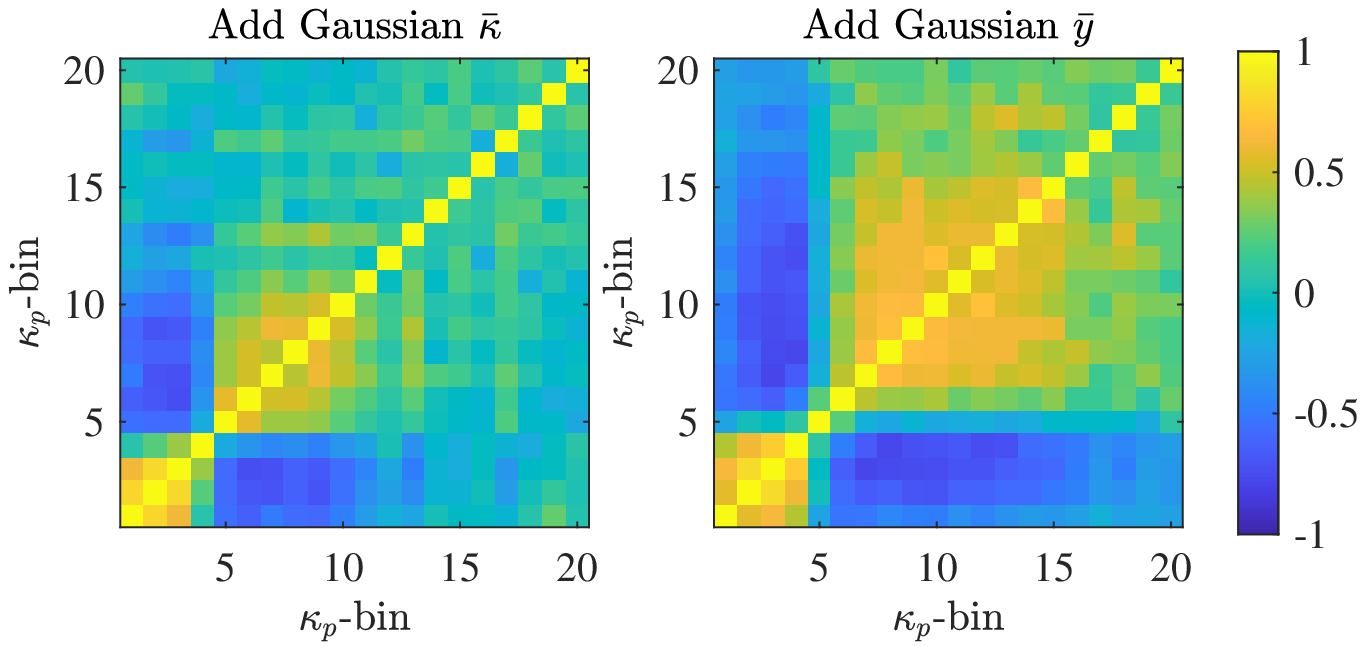}\\
\includegraphics[width=8.5cm]{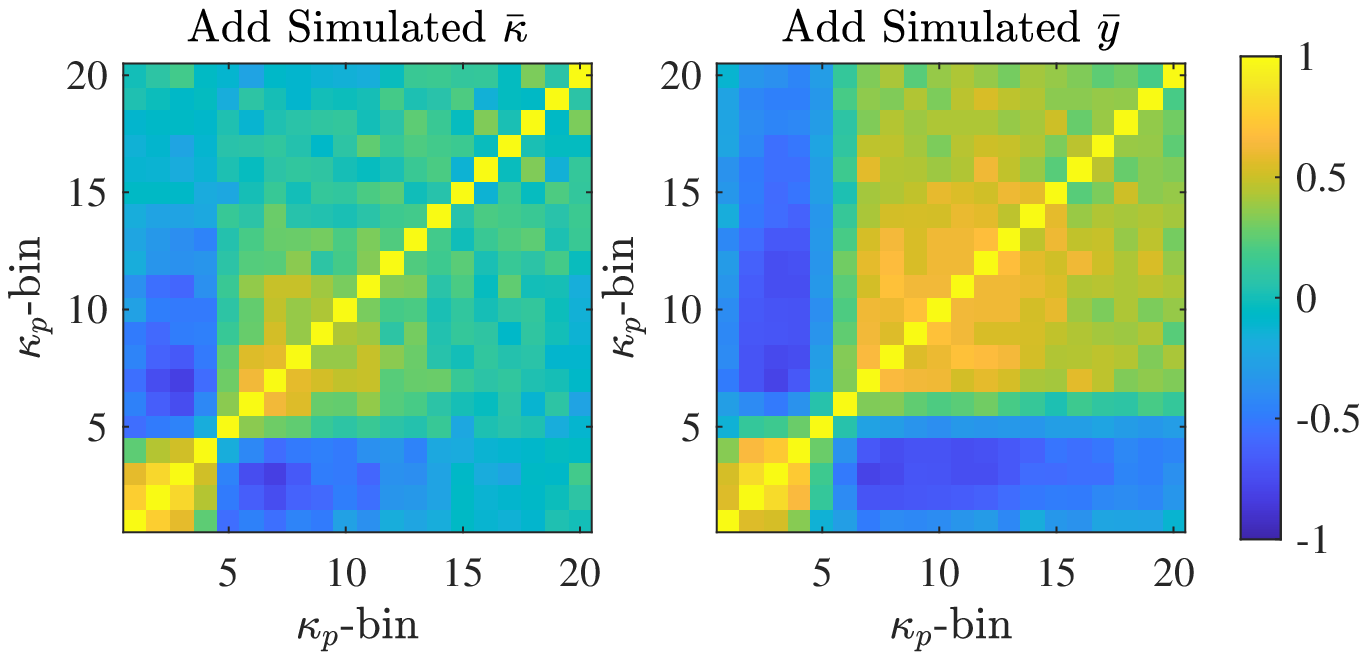}
\caption{ The correlation matrices improved by method i - iv (detailed in Section \ref{subsec:ssv}) are presented in the top left, top right, bottom left and bottom right panel, respectively.
The top two cases assume that the distribution of the mean $\kappa$/$y$ is Gaussian, and the bottom two methods directly add simulated mean value into corresponding mocks.
We can find that adding mean value on $y$ maps pre to inverse-transform recover stronger features for the off-diagonal matrix elements than directly modifying $\kappa$ maps.
Comparing with Fig.\ref{fig:covariance}, the method ii produces the most similar pattern to the simulated peak height counts covariance. }
\label{fig:change}
\end{figure}

By construction, the $y$ maps generated with the given power spectrum have the mean of zero.
This is shown in the bottom right panel of Fig. \ref{fig:mean}.
The PDF of the $\bar{y}$ is a delta function at zero.
While the Gaussianized maps have a broad scatter of the mean, which is shown in the upper right panel, with $\sigma(\bar{y})=0.1327$.
After the inverse-transform, we find that the mock $\bar\kappa$ is not exactly zero (bottom left panel), but the distribution is still very narrow, which is very different from the simulated maps (upper left panel).

The lack of the variance in the mean value comes from the lack of super sample variance, i.e. the long wave perturbation beyond the map size we investigated.
The mock convergence maps have no power beyond the map size by design.
In simulation, the light-cone is cut off from the stacked boxes.
Although the simulated maps also miss some super sample variance beyond the simulation box, they contain the long wavelength perturbation beyond the map size and the mean value of the 96 realizations has a scatter $\sigma(\bar\kappa)=0.0020$.
 
Here we do not try to add the exact super sample variance corresponding to the light-cone construction.
We mimic the super sample variance effect on each map by several approaches.
\bi
\item[ i ] \emph{ Add Gaussian $\bar\kappa$. }
For each $\kappa$ mock inverse-transformed from Gaussian random fields, we add a random number drawn from a Gaussian distribution with zero mean and scatter measured from the simulated convergence maps. This has no impact on the power spectrum result. The normalized peak height counts covariance is presented in the top left panel of Fig. \ref{fig:change}.
\item[ ii ] \emph{ Add Gaussian $\bar{y}$. }
For each $y$ map we added a random number drawn from a Gaussian distribution with the scatter same as the Gaussianized $y$ fields. Then we transform the new $y$ fields inversely with the $\kappa$--$y$ relations and obtain the new $\kappa$ realizations. These new $\kappa$ fields have the scatter of the mean similar to the simulated ones. The power spectrum only shows negligible change and we do not present the result here. The normalized peak height counts covariance is presented in the top right panel of Fig. \ref{fig:change}.
\item[ iii ] \emph{ Add simulated $\bar\kappa$. }
We add the mean values taken from the simulated convergence maps to our mock $\kappa$ maps. This process does not assume the distribution of the mean is Gaussian, but the mocks have the exact $\bar\kappa$ distribution with the simulated ones. The result is presented in the bottom left panel of Fig. \ref{fig:change}.
\item[ iv ] \emph{ Add simulated $\bar{y}$. }
We add the $\bar{y}$'s taken from the Gaussianized convergence maps to our mock $y$ maps.  Same as the previous one, the $\bar{y}$ distribution is exactly the same with the Gaussianized fields. The result is shown in the bottom right panel of Fig. \ref{fig:change}.
\ei
All of these approaches can recover the features in the simulated peak height counts covariance to some extent.
Careful readers can find that the strong correlations between the large $\kappa_\subp$ bins, between the small $\kappa_\subp$ bins, and the anti-correlation between large and small $\kappa_\subp$ bins are not perfectly reproduced.
Compared to the simulated case, the (anti-)correlation is weaker for method i and iii.
Method ii and iv produces stronger (anti-)correlation.
The method ii produces the most similar result with the simulated one.
These results tell us that the features in the simulated one are mainly from the super sample variance, and this effect can be reproduced by adding reasonable scatters in the mean value of these maps.




\begin{figure}[ht!]
\centering
\includegraphics[width=8.5cm]{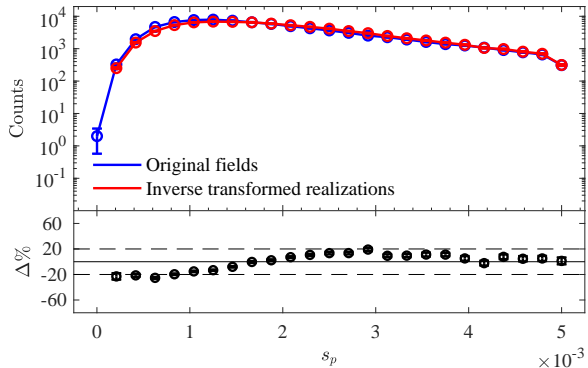}
\caption{ Similar to Fig. \ref{fig:peak}, the distributions of peak steepness are shown in this plot. The blue line represents steepness counts of original simulated $\kappa$ maps, while the result of mocks by the inverse-Gaussianization method is shown by the red line. 
We find that the steepness counts are well produced for almost all the range we consider. }
\label{fig:steepness}
\end{figure}

\subsection{ Peak Steepness }
\label{subsec:steepness}

\citealt{Ribli2018Learning} proposed an improved lensing observable motivated by deep learning, which is called peak steepness. 
The peak steepness counts perform better than the peak height counts on constraining $\sigma_8$ and $\Omega_m$.
So we also test the peak steepness counts and its covariance on the mocks.

Peak steepness means the difference of peaks and their neighbors. Naturally, the gradient at the peak position must be zero.
As in \citealt{Ribli2018Learning}, we describe the steepness of peaks by the magnitudes of gradients near the peaks, which can be calculated by the convolution with the isotropic discrete Laplace operator. The operator we use has the following form:
\begin{equation}
L = 4\left[\begin{array}{ccc}{0} & {-0.25} & {0} \\ {-0.25} & {1} & {-0.25} \\ {0} & {-0.25} & {0}\end{array}\right]\ .
\label{eq:laplace}
\end{equation}
We use the notation $s_\subp$ to denote the peak steepness, and by design $s_\subp>0$.

Similar to Fig. \ref{fig:peak}, we present the peak steepness counts $N(s_\subp)$ in Fig. \ref{fig:steepness}.
The blue line is the result of simulated lensing convergence maps and the red one is from the inverse-Gaussianization method.
We find that the difference is within $20\%$ for almost all the steepness range we investigated.
An overall trend is that the number of low steepness peaks ($s_p \lesssim 1.7\times 10^{-3}$) is smaller by $0-20\%$ than the simulation and the high steepness peaks ($2\times 10^{-3}<s_p <4\times 10^{-3}$) are overproduced by similar amounts.
Notice that the steepness is sensitive to the pixel size and smoothing scale.
To our knowledge, the dependence of the steepness measure on these factors has not been investigated in detail and it is out of the scope of this work.
We expect that increasing the smoothing scale and pixel size will affect the shape of the local transform function, but will not affect the performance of the inverse-Gaussianization much.
Our study may give some clues for modeling of the steepness statistic.


\subsection{Peak Steepness Covariance}
\label{subsec:steepnesscov}

Similar to peak height counts, we also study the covariance of peak steepness counts which is defined as $\cov_{i j}=\left\langle \Delta N\left(s_{\mathrm{p}, i}\right) \Delta  N\left(s_{\mathrm{p}, j}\right)\right\rangle$. 
We divide the peaks into 20 $s_\subp$-bins linearly covering $3\times10^{-4}<s_\subp<5\times10^{-3}$.

\begin{figure}[ht!]
\centering
\includegraphics[width=8.5cm]{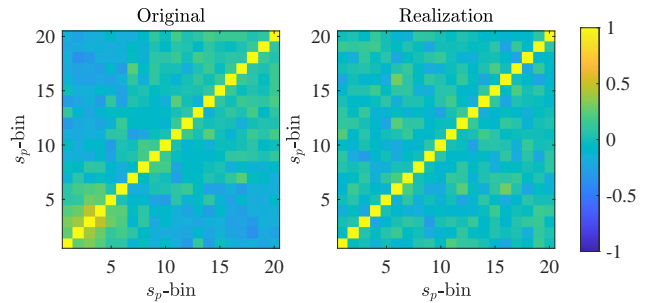}
\caption{ The normalized covariance for peak steepness counts of the simulated and mock maps is presented in the left and right panel, respectively.
The result from simulations has some weak off-diagonal elements while the matrix of mocks is diagonal. Both of matrix have the same steepness value range at $[3\times10^{-4},5\times10^{-3}]$. This result indicates the absence of super sample variance.}
\label{fig:pic5_steepness}
\end{figure}

The weak correlation at different steepness in the simulated maps is shown in the left panel of Fig. \ref{fig:pic5_steepness}. The right panel is the normalized steepness covariance of 96 mocks by the inverse-Gaussianization method.
It is approximately a diagonal matrix. This situation is similar to the peak height counts covariance.
The missing feature in the covariance is caused by the lack of super sample variance.

However, different from the peak height, peak steepness is immune to a constant shift by definition.
Thus, we apply the method ii and iv described in Section \ref{subsec:ssv} to recover the features in the steepness covariance. 
We find that both methods recover the features but overproduce the correlation and anti-correlation among $s_\subp$-bins.

\begin{figure}[ht!]
\centering
\includegraphics[width=8.5cm]{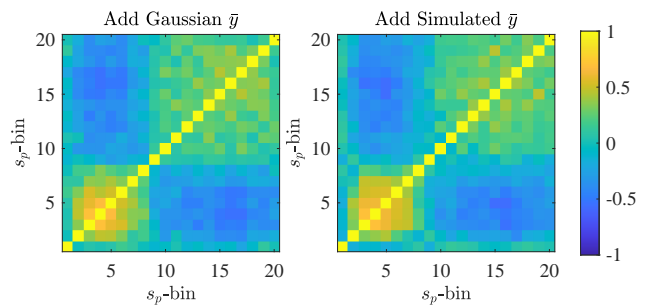}
\caption{ 
 We mimic the super sample effect by two of the four methods detailed in Sec. \ref{subsec:ssv}. 
Identical to Fig. \ref{fig:pic5_steepness}, the result of method ii is shown in the left panel, while the right normalized covariance matrix is calculated by the last method in Sec. \ref{subsec:ssv}. We find that both methods produce the matrices with stronger (anti-)correlation.  This result is consistent with the result of peak height counts covariance. 
}
\label{fig:pic10}
\end{figure}

\section{Application}
\label{sec4}

Joint analysis of the two-point statistics and the high-order one such as peak height counts can break the degeneracy in the cosmological parameter constraints.
However, they are not independent observables.
The precision of the covariance between these observables is essential in this joint analysis to avoid the mis-estimation of the constraints.
A large number of mocks are required to quantify the covariance, which is out of the capability if the full-resolution simulation is used.
In this section, we will focus on an important application of the inverse-Gaussianization method, calculating the covariance matrix between the power spectrum and peak statistics.

Assume the cosmological parameter we care is $\lambda$ and the observables are band power $C(\ell_i)$ with $i=1,\cdots,m$ and peak height counts $N(\kappa_{\mathrm{p},j})$ with $j=1,\cdots,n$.  
The data vector $\datap=[C(\ell_i),N(\kappa_{\mathrm{p},j})]$.
The likelihood of the joint analysis is
\begin{equation}
\ln\mathcal{L}\propto 
\frac{\partial \ln\datap}{\partial \lambda}
\hat\cov^{-1}
\frac{\partial \ln\datap^T}{\partial \lambda} \ .
\end{equation}
Here the normalized joint covariance $\hat\cov$ has a size of $(m+n)\times(m+n)$ and contains the correlation between band power and peak counts on different bins, and the correlation between the two.

\begin{figure}[ht!]
\centering
\includegraphics[width=8.5cm]{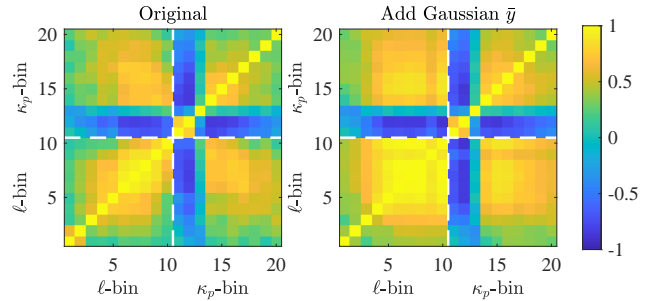}
\caption{
Both maps represent the covariance matrix between the power spectrum and peak height counts. The 1-10 bins in both the x-axis and y-axis correspond with the power spectrum, while the 11-20 linear bins are the peak height counts for $\kappa_p \in [-0.02,0.1]$. The left panel is calculated by 96 original simulation fields, while we obtain the right map from 40,000 mocks generated by our improved inverse-Gaussianization method. We found that there is little difference between the two maps which is caused by the numbers of data. The matrix from mocks is more accurate than the left matrix, especially when compared with Fig. \ref{fig:covariance}. }
\label{fig:ps_pc}
\end{figure}

We set the number of bins to be $10$ for both the power spectrum and peak height counts.
So the size of full covariance is $20\times20$. 
In order to obtain the precise covariance matrix, we use 40,000 weak gravitational lensing mocks generated by the inverse-Gaussianization method with the inclusion of the scatter of $\bar{y}$ in a Gaussian form (method ii).
The result is presented in Fig. \ref{fig:ps_pc}.
The left plot is the result from 96 simulated maps,
while the right plot is the result of 40,000 mocks.
The joint covariance from the mocks is obviously more smooth and less noisy.
However, the inverse-Gaussianization method overproduces the correlation among the peak height bins and the cross-correlation between the two observables.
We argue that this discrepancy comes from the inaccuracy in adding the super-sample effects.

\begin{figure}[ht!]
\centering
\includegraphics[width=8.5cm]{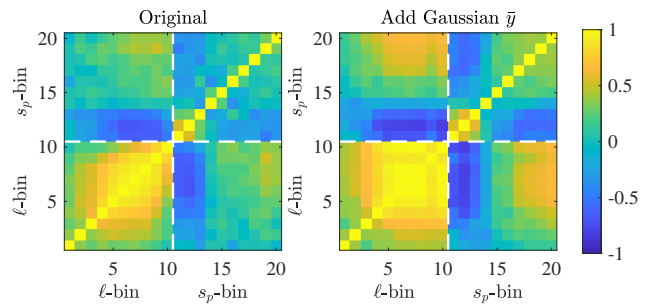}
\caption{
Identical to Fig.\ref{fig:ps_pc}, but for the covariance matrix between power spectrum and peak steepness counts. The last 10 bins of both x-axis and y-axis represent the value of convergence peak steepness covering the range for $3\times10^{-4}<s_p<5\times10^{-3}$.}
\label{fig:ps_sc}
\end{figure}

We also present the joint covariance for power spectrum and peak steepness from the same 40,000 convergence mocks in the right plot in Fig. \ref{fig:ps_sc}
and the left plot is the result of simulated maps. 
Compared to the simulated one, our method overestimate the covariance, and again, this discrepancy comes from the difference in the super sample variance in two cases.
In particular, we find the correlation between the power spectrum and peak steepness counts is significantly weaker than the joint of the power spectrum and peak height counts.
Based on the findings in \citealt{Ribli2018Learning}, the peak steepness is more powerful than peak height on the constraints of $\Omega_m$ and $\sigma_8$. The weaker correlation between the two observables means that they provide more independent information.
The joint of second order statistics and peak steepness has great potential to break the degeneracy of $S_8$.

\section{Conclusion and Discussion}
\label{sec5}

The next stage weak lensing survey has great statistical power, implying the ability to use non-Gaussian statistics to break the cosmological parameter degeneracy in the joint analysis.
In this work, we focused on the peak statistics and investigated whether the peak statistics could be reproduced by the inverse-Gaussianization method.

We used 96 simulated weak lensing convergence maps at $z_s=1$ to obtain the Gaussianization transform function, and the power spectrum of the Gaussianized fields.
We fast generated the lensing mocks by generating the Gaussian random fields according to the above power spectrum and then inverse transform them.
The resulting mocks have the same power spectrum with the simulated one.
The peak height counts are also well reproduced.
The difference from the simulated one is $<20\%$ for $-0.015<\kappa_p<0.065$ ($\kappa_p/\sigma_{\kappa}\in [-1.2, 5.4]$) which covers the common value range of the recent weak lensing peak research.
The peak steepness counts are also well reproduced and the difference from the simulation is $<20\%$ for almost all the steepness range we investigated.

The covariance for the peak height counts and peak steepness counts is very close to diagonal, in contrast with the simulated cases.
We found that the lack of super sample variance is the main reason for this difference.
We proposed several methods to mimic the super sample effects and the (anti-)correlation among different bins is recovered.
The remaining difference comes from the fact that we only mimic the super sample variance by tuning the mean value for these mocks.

We used this improved inverse-Gaussianization method with the inclusion of effective sample variance to obtain the covariance for the joint analysis of the power spectrum and the peak statistics. 
A greater potential of the combination between power spectrum and peak steepness was found, compared to the joint analysis between power spectrum and peak heights. 

The inverse-Gaussianization method is to fast generate noiseless convergence fields based on a small patch of weak lensing simulations.
They could be converted to cosmic shear fields, and further include the noise and mask effect to mimic the observation.

Our method produces slightly better distribution for the peak height counts than the result in \citealt{2017MNRAS.465.1974S}.
Both methods use the local transform to obtain lensing mocks and they differ in details.
The main difference is that
\citealt{2017MNRAS.465.1974S} uses theoretical power spectrum with corrections for resolution effects to obtain the power spectrum after Gaussianization.
The new Gaussian random $y$ fields are produced by this power spectrum with continuous $\ell$.
Our method directly measures the power spectrum from these Gaussianized fields, and the new $y$ fields are produced from the measured band power.
This process automatically includes the numerical/resolution effects.

The following details worth further investigation in the future.
The missing of the rare extremely high peaks in the inverse-Gaussianization method could be partially solved by considering the scatter in the local transform function especially at the high end.
However, these rare highest peaks correspond to very nonlinear small scale structures where the Gaussianization performs worse.
It is interesting to study how well the inverse-Gaussianization method with further improvement could capture this complicated information.

The direct inverse-Gaussianization method can not produce strong features in peak statistics covariance.
This is due to the lack of super sample variance.
Note that the simulated maps we use also miss some super sample modes since it is cut out from a light-cone constructed by stacking cubic simulation boxes.
In principle, we could increase the map size to include the super sample variance.
In the case that we do not have such a weak lensing simulation covering a very large portion of the sky, it is still doable thanks to the linearity at large scales.
Given the cosmological parameters, the power spectrum at large scales is well predicted.
Also at these scales, the power spectrum of the Gaussianized field is close to the power spectrum of the convergence field.
We could adopt the theoretically predicted large scale power to construct even the full-sky weak lensing map.

\section{acknowledgments}

We thank Pengjie Zhang, Jun Zhang, Shuo Yuan for useful discussions.
This work was supported by the National Key Basic Research and Development Program of China (No. 2018YFA0404504), and the National Science Foundation of China (grants No. 11773048, 11621303, 11890691). X.K.L. acknowledges the support from NSFC-11803028 and YNU Grant C176220100008.


\bibliographystyle{aasjournal}
\bibliography{paperNotes.bib}

\begin{thebibliography}{}
\expandafter\ifx\csname natexlab\endcsname\relax\def\natexlab#1{#1}\fi
\providecommand{\url}[1]{\href{#1}{#1}}

\bibitem[{Abbott {et~al.}(2018)Abbott, Abdalla, Aleksić, Allam, Amara, Bacon,
  Balbinot, Banerji, Bechtol, \& Benoit-Lévy}]{Abbott2018The}
Abbott, T., Abdalla, F.~B., Aleksić, J., {et~al.} 2018, Monthly Notices of the
  Royal Astronomical Society, 460, 1270

\bibitem[{{Abbott} {et~al.}(2018){Abbott}, {Abdalla}, {Alarcon}, {Aleksi{\'c}},
  {Allam}, {Allen}, {Amara}, {Annis}, {Asorey}, {Avila}, {Bacon}, {Balbinot},
  {Banerji}, {Banik}, {Barkhouse}, {Baumer}, {Baxter}, {Bechtol}, {Becker},
  {Benoit-L{\'e}vy}, {Benson}, {Bernstein}, {Bertin}, {Blazek}, {Bridle},
  {Brooks}, {Brout}, {Buckley-Geer}, {Burke}, {Busha}, {Campos}, {Capozzi},
  {Carnero Rosell}, {Carrasco Kind}, {Carretero}, {Castander}, {Cawthon},
  {Chang}, {Chen}, {Childress}, {Choi}, {Conselice}, {Crittenden}, {Crocce},
  {Cunha}, {D'Andrea}, {da Costa}, {Das}, {Davis}, {Davis}, {De Vicente},
  {DePoy}, {DeRose}, {Desai}, {Diehl}, {Dietrich}, {Dodelson}, {Doel},
  {Drlica-Wagner}, {Eifler}, {Elliott}, {Elsner}, {Elvin-Poole}, {Estrada},
  {Evrard}, {Fang}, {Fernandez}, {Fert{\'e}}, {Finley}, {Flaugher}, {Fosalba},
  {Friedrich}, {Frieman}, {Garc{\'\i}a-Bellido}, {Garcia-Fernandez}, {Gatti},
  {Gaztanaga}, {Gerdes}, {Giannantonio}, {Gill}, {Glazebrook}, {Goldstein},
  {Gruen}, {Gruendl}, {Gschwend}, {Gutierrez}, {Hamilton}, {Hartley}, {Hinton},
  {Honscheid}, {Hoyle}, {Huterer}, {Jain}, {James}, {Jarvis}, {Jeltema},
  {Johnson}, {Johnson}, {Kacprzak}, {Kent}, {Kim}, {King}, {Kirk}, {Kokron},
  {Kovacs}, {Krause}, {Krawiec}, {Kremin}, {Kuehn}, {Kuhlmann}, {Kuropatkin},
  {Lacasa}, {Lahav}, {Li}, {Liddle}, {Lidman}, {Lima}, {Lin}, {MacCrann},
  {Maia}, {Makler}, {Manera}, {March}, {Marshall}, {Martini}, {McMahon},
  {Melchior}, {Menanteau}, {Miquel}, {Miranda}, {Mudd}, {Muir}, {M{\"o}ller},
  {Neilsen}, {Nichol}, {Nord}, {Nugent}, {Ogando}, {Palmese}, {Peacock},
  {Peiris}, {Peoples}, {Percival}, {Petravick}, {Plazas}, {Porredon}, {Prat},
  {Pujol}, {Rau}, {Refregier}, {Ricker}, {Roe}, {Rollins}, {Romer}, {Roodman},
  {Rosenfeld}, {Ross}, {Rozo}, {Rykoff}, {Sako}, {Salvador}, {Samuroff},
  {S{\'a}nchez}, {Sanchez}, {Santiago}, {Scarpine}, {Schindler}, {Scolnic},
  {Secco}, {Serrano}, {Sevilla-Noarbe}, {Sheldon}, {Smith}, {Smith}, {Smith},
  {Soares-Santos}, {Sobreira}, {Suchyta}, {Tarle}, {Thomas}, {Troxel},
  {Tucker}, {Tucker}, {Uddin}, {Varga}, {Vielzeuf}, {Vikram}, {Vivas},
  {Walker}, {Wang}, {Wechsler}, {Weller}, {Wester}, {Wolf}, {Yanny}, {Yuan},
  {Zenteno}, {Zhang}, {Zhang}, {Zuntz}, \& {Dark Energy Survey
  Collaboration}}]{2018PhRvD..98d3526A}
{Abbott}, T.~M.~C., {Abdalla}, F.~B., {Alarcon}, A., {et~al.} 2018, \prd, 98,
  043526

\bibitem[{{Agrawal} {et~al.}(2017){Agrawal}, {Makiya}, {Chiang}, {Jeong},
  {Saito}, \& {Komatsu}}]{2017JCAP...10..003A}
{Agrawal}, A., {Makiya}, R., {Chiang}, C.-T., {et~al.} 2017, \jcap, 2017, 003

\bibitem[{{Aihara} {et~al.}(2018){Aihara}, {Armstrong}, {Bickerton}, {Bosch},
  {Coupon}, {Furusawa}, {Hayashi}, {Ikeda}, {Kamata}, {Karoji}, {Kawanomoto},
  {Koike}, {Komiyama}, {Lang}, {Lupton}, {Mineo}, {Miyatake}, {Miyazaki},
  {Morokuma}, {Obuchi}, {Oishi}, {Okura}, {Price}, {Takata}, {Tanaka},
  {Tanaka}, {Tanaka}, {Uchida}, {Uraguchi}, {Utsumi}, {Wang}, {Yamada},
  {Yamanoi}, {Yasuda}, {Arimoto}, {Chiba}, {Finet}, {Fujimori}, {Fujimoto},
  {Furusawa}, {Goto}, {Goulding}, {Gunn}, {Harikane}, {Hattori}, {Hayashi},
  {He{\l}miniak}, {Higuchi}, {Hikage}, {Ho}, {Hsieh}, {Huang}, {Huang},
  {Imanishi}, {Iwata}, {Jaelani}, {Jian}, {Kashikawa}, {Katayama}, {Kojima},
  {Konno}, {Koshida}, {Kusakabe}, {Leauthaud}, {Lee}, {Lin}, {Lin},
  {Mandelbaum}, {Matsuoka}, {Medezinski}, {Miyama}, {Momose}, {More}, {More},
  {Mukae}, {Murata}, {Murayama}, {Nagao}, {Nakata}, {Niida}, {Niikura},
  {Nishizawa}, {Oguri}, {Okabe}, {Ono}, {Onodera}, {Onoue}, {Ouchi}, {Pyo},
  {Shibuya}, {Shimasaku}, {Simet}, {Speagle}, {Spergel}, {Strauss}, {Sugahara},
  {Sugiyama}, {Suto}, {Suzuki}, {Tait}, {Takada}, {Terai}, {Toba}, {Turner},
  {Uchiyama}, {Umetsu}, {Urata}, {Usuda}, {Yeh}, \&
  {Yuma}}]{2018PASJ...70S...8A}
{Aihara}, H., {Armstrong}, R., {Bickerton}, S., {et~al.} 2018, \pasj, 70, S8

\bibitem[{{Bartelmann} \& {Schneider}(2001)}]{2001PhR...340..291B}
{Bartelmann}, M., \& {Schneider}, P. 2001, \physrep, 340, 291

\bibitem[{{Blot} {et~al.}(2019){Blot}, {Crocce}, {Sefusatti}, {Lippich},
  {S{\'a}nchez}, {Colavincenzo}, {Monaco}, {Alvarez}, {Agrawal}, {Avila},
  {Balaguera-Antol{\'\i}nez}, {Bond}, {Codis}, {Dalla Vecchia}, {Dorta},
  {Fosalba}, {Izard}, {Kitaura}, {Pellejero-Ibanez}, {Stein}, {Vakili}, \&
  {Yepes}}]{2019MNRAS.485.2806B}
{Blot}, L., {Crocce}, M., {Sefusatti}, E., {et~al.} 2019, \mnras, 485, 2806

\bibitem[{{Chuang} {et~al.}(2015){Chuang}, {Kitaura}, {Prada}, {Zhao}, \&
  {Yepes}}]{2015MNRAS.446.2621C}
{Chuang}, C.-H., {Kitaura}, F.-S., {Prada}, F., {Zhao}, C., \& {Yepes}, G.
  2015, \mnras, 446, 2621

\bibitem[{{Colavincenzo} {et~al.}(2019){Colavincenzo}, {Sefusatti}, {Monaco},
  {Blot}, {Crocce}, {Lippich}, {S{\'a}nchez}, {Alvarez}, {Agrawal}, {Avila},
  {Balaguera-Antol{\'\i}nez}, {Bond}, {Codis}, {Dalla Vecchia}, {Dorta},
  {Fosalba}, {Izard}, {Kitaura}, {Pellejero-Ibanez}, {Stein}, {Vakili}, \&
  {Yepes}}]{2019MNRAS.482.4883C}
{Colavincenzo}, M., {Sefusatti}, E., {Monaco}, P., {et~al.} 2019, \mnras, 482,
  4883

\bibitem[{{Coulton} {et~al.}(2019){Coulton}, {Liu}, {Madhavacheril},
  {B{\"o}hm}, \& {Spergel}}]{2019JCAP...05..043C}
{Coulton}, W.~R., {Liu}, J., {Madhavacheril}, M.~S., {B{\"o}hm}, V., \&
  {Spergel}, D.~N. 2019, \jcap, 2019, 043

\bibitem[{De~Jong {et~al.}(2015)De~Jong, Kleijn, Boxhoorn, Buddelmeijer,
  Capaccioli, Getman, Grado, Helmich, Huang, \& Irisarri}]{De2015The}
De~Jong, J. T.~A., Kleijn, G. A.~V., Boxhoorn, D.~R., {et~al.} 2015, Astronomy
  \& Astrophysics, 582, A62

\bibitem[{{Dietrich} \& {Hartlap}(2010)}]{2010MNRAS.402.1049D}
{Dietrich}, J.~P., \& {Hartlap}, J. 2010, \mnras, 402, 1049

\bibitem[{{Dodelson} \& {Schneider}(2013)}]{2013PhRvD..88f3537D}
{Dodelson}, S., \& {Schneider}, M.~D. 2013, \prd, 88, 063537

\bibitem[{{Fan} {et~al.}(2010){Fan}, {Shan}, \& {Liu}}]{2010ApJ...719.1408F}
{Fan}, Z., {Shan}, H., \& {Liu}, J. 2010, \apj, 719, 1408

\bibitem[{{Fu} {et~al.}(2008){Fu}, {Semboloni}, {Hoekstra}, {Kilbinger}, {van
  Waerbeke}, {Tereno}, {Mellier}, {Heymans}, {Coupon}, {Benabed}, {Benjamin},
  {Bertin}, {Dor{\'e}}, {Hudson}, {Ilbert}, {Maoli}, {Marmo}, {McCracken}, \&
  {M{\'e}nard}}]{2008A&A...479....9F}
{Fu}, L., {Semboloni}, E., {Hoekstra}, H., {et~al.} 2008, \aap, 479, 9

\bibitem[{{Fu} {et~al.}(2014){Fu}, {Kilbinger}, {Erben}, {Heymans},
  {Hildebrandt}, {Hoekstra}, {Kitching}, {Mellier}, {Miller}, {Semboloni},
  {Simon}, {Van Waerbeke}, {Coupon}, {Harnois-D{\'e}raps}, {Hudson}, {Kuijken},
  {Rowe}, {Schrabback}, {Vafaei}, \& {Velander}}]{2014MNRAS.441.2725F}
{Fu}, L., {Kilbinger}, M., {Erben}, T., {et~al.} 2014, \mnras, 441, 2725

\bibitem[{{Gatti} {et~al.}(2019){Gatti}, {Chang}, {Friedrich}, {Jain}, {Bacon},
  {Crocce}, {DeRose}, {Ferrero}, {Fosalba}, {Gaztanaga}, {Gruen}, {Harrison},
  {Jeffrey}, {MacCrann}, {McClintock}, {Secco}, {Whiteway}, {Abbott}, {Allam},
  {Annis}, {Avila}, {Brooks}, {Buckley-Geer}, {Burke}, {Carnero Rosell},
  {Carrasco Kind}, {Carretero}, {Cawthon}, {Crocce}, {da Costa}, {De Vicente},
  {Desai}, {Diehl}, {Doel}, {Eifler}, {Estrada}, {Everett}, {Evrard},
  {Fosalba}, {Frieman}, {Garcia-Bellido}, {Gerdes}, {Gruendl}, {Gschwend},
  {Gutierrez}, {James}, {Johnson}, {Krause}, {Kuehn}, {Kuehn}, {Lima}, {Maia},
  {March}, {Marshall}, {Melchior}, {Menanteau}, {Miquel}, {Palmese},
  {Paz-Chinchon}, {Plazas}, {Sanchez}, {Sanchez}, {Scarpine}, {Schubnell},
  {Serrano}, {Sevilla-Noarbe}, {Smith}, {Soares-Santos}, {Suchyta}, {Swanson},
  {Tarle}, {Thomas}, {Troxel}, {Zuntz}, \& {the DES
  collaboration}}]{2019arXiv191105568G}
{Gatti}, M., {Chang}, C., {Friedrich}, O., {et~al.} 2019, arXiv e-prints,
  arXiv:1911.05568

\bibitem[{{Giocoli} {et~al.}(2018){Giocoli}, {Moscardini}, {Baldi},
  {Meneghetti}, \& {Metcalf}}]{2018MNRAS.478.5436G}
{Giocoli}, C., {Moscardini}, L., {Baldi}, M., {Meneghetti}, M., \& {Metcalf},
  R.~B. 2018, \mnras, 478, 5436

\bibitem[{{Hall} \& {Mead}(2018)}]{2018MNRAS.473.3190H}
{Hall}, A., \& {Mead}, A. 2018, \mnras, 473, 3190

\bibitem[{{Hamana} {et~al.}(2004){Hamana}, {Takada}, \&
  {Yoshida}}]{Hamana2010Searching}
{Hamana}, T., {Takada}, M., \& {Yoshida}, N. 2004, \mnras, 350, 893

\bibitem[{{Hildebrandt}(2014)}]{2014arXiv1406.1379H}
{Hildebrandt}, H. 2014, arXiv e-prints, arXiv:1406.1379

\bibitem[{{Hildebrandt} {et~al.}(2017){Hildebrandt}, {Viola}, {Heymans},
  {Joudaki}, {Kuijken}, {Blake}, {Erben}, {Joachimi}, {Klaes}, {Miller},
  {Morrison}, {Nakajima}, {Verdoes Kleijn}, {Amon}, {Choi}, {Covone}, {de
  Jong}, {Dvornik}, {Fenech Conti}, {Grado}, {Harnois-D{\'e}raps}, {Herbonnet},
  {Hoekstra}, {K{\"o}hlinger}, {McFarland}, {Mead}, {Merten}, {Napolitano},
  {Peacock}, {Radovich}, {Schneider}, {Simon}, {Valentijn}, {van den Busch},
  {van Uitert}, \& {Van Waerbeke}}]{2017MNRAS.465.1454H}
{Hildebrandt}, H., {Viola}, M., {Heymans}, C., {et~al.} 2017, \mnras, 465, 1454

\bibitem[{{Izard} {et~al.}(2018){Izard}, {Fosalba}, \&
  {Crocce}}]{2018MNRAS.473.3051I}
{Izard}, A., {Fosalba}, P., \& {Crocce}, M. 2018, \mnras, 473, 3051

\bibitem[{{Jarvis} {et~al.}(2004){Jarvis}, {Bernstein}, \&
  {Jain}}]{2004MNRAS.352..338J}
{Jarvis}, M., {Bernstein}, G., \& {Jain}, B. 2004, \mnras, 352, 338

\bibitem[{Joachimi {et~al.}(2011)Joachimi, Taylor, \&
  Kiessling}]{Joachimi2011Cosmological}
Joachimi, B., Taylor, A.~N., \& Kiessling, A. 2011, Monthly Notices of the
  Royal Astronomical Society, 418, 145

\bibitem[{{Kacprzak} {et~al.}(2016){Kacprzak}, {Kirk}, {Friedrich}, {Amara},
  {Refregier}, {Marian}, {Dietrich}, {Suchyta}, {Aleksi{\'c}}, {Bacon},
  {Becker}, {Bonnett}, {Bridle}, {Chang}, {Eifler}, {Hartley}, {Huff},
  {Krause}, {MacCrann}, {Melchior}, {Nicola}, {Samuroff}, {Sheldon}, {Troxel},
  {Weller}, {Zuntz}, {Abbott}, {Abdalla}, {Armstrong}, {Benoit-L{\'e}vy},
  {Bernstein}, {Bernstein}, {Bertin}, {Brooks}, {Burke}, {Carnero Rosell},
  {Carrasco Kind}, {Carretero}, {Castander}, {Crocce}, {D'Andrea}, {da Costa},
  {Desai}, {Diehl}, {Evrard}, {Neto}, {Flaugher}, {Fosalba}, {Frieman},
  {Gerdes}, {Goldstein}, {Gruen}, {Gruendl}, {Gutierrez}, {Honscheid}, {Jain},
  {James}, {Jarvis}, {Kuehn}, {Kuropatkin}, {Lahav}, {Lima}, {March},
  {Marshall}, {Martini}, {Miller}, {Miquel}, {Mohr}, {Nichol}, {Nord},
  {Plazas}, {Romer}, {Roodman}, {Rykoff}, {Sanchez}, {Scarpine}, {Schubnell},
  {Sevilla-Noarbe}, {Smith}, {Soares-Santos}, {Sobreira}, {Swanson}, {Tarle},
  {Thomas}, {Vikram}, {Walker}, {Zhang}, \& {DES
  Collaboration}}]{2016MNRAS.463.3653K}
{Kacprzak}, T., {Kirk}, D., {Friedrich}, O., {et~al.} 2016, \mnras, 463, 3653

\bibitem[{Kilbinger(2015)}]{Kilbinger2015Cosmology}
Kilbinger, M. 2015, Reports on Progress in Physics, 78, 086901

\bibitem[{{K{\"o}hlinger} {et~al.}(2017){K{\"o}hlinger}, {Viola}, {Joachimi},
  {Hoekstra}, {van Uitert}, {Hildebrandt}, {Choi}, {Erben}, {Heymans},
  {Joudaki}, {Klaes}, {Kuijken}, {Merten}, {Miller}, {Schneider}, \&
  {Valentijn}}]{2017MNRAS.471.4412K}
{K{\"o}hlinger}, F., {Viola}, M., {Joachimi}, B., {et~al.} 2017, \mnras, 471,
  4412

\bibitem[{{Kratochvil} {et~al.}(2010){Kratochvil}, {Haiman}, \&
  {May}}]{2010PhRvD..81d3519K}
{Kratochvil}, J.~M., {Haiman}, Z., \& {May}, M. 2010, \prd, 81, 043519

\bibitem[{Laureijs {et~al.}(2011)Laureijs, Amiaux, Arduini, Auguères,
  Brinchmann, Cole, Cropper, Dabin, Duvet, \& Ealet}]{Laureijs2011Euclid}
Laureijs, R., Amiaux, J., Arduini, S., {et~al.} 2011, HAL - OBSPM

\bibitem[{Lee \& Pen(2008)}]{Lee2008Information}
Lee, J., \& Pen, U.~L. 2008, Astrophysical Journal, 686, L1

\bibitem[{{Lin} \& {Kilbinger}(2015{\natexlab{a}})}]{2015A&A...576A..24L}
{Lin}, C.-A., \& {Kilbinger}, M. 2015{\natexlab{a}}, \aap, 576, A24

\bibitem[{{Lin} \& {Kilbinger}(2015{\natexlab{b}})}]{2015A&A...583A..70L}
---. 2015{\natexlab{b}}, \aap, 583, A70

\bibitem[{{Lippich} {et~al.}(2019){Lippich}, {S{\'a}nchez}, {Colavincenzo},
  {Sefusatti}, {Monaco}, {Blot}, {Crocce}, {Alvarez}, {Agrawal}, {Avila},
  {Balaguera-Antol{\'\i}nez}, {Bond}, {Codis}, {Dalla Vecchia}, {Dorta},
  {Fosalba}, {Izard}, {Kitaura}, {Pellejero-Ibanez}, {Stein}, {Vakili}, \&
  {Yepes}}]{2019MNRAS.482.1786L}
{Lippich}, M., {S{\'a}nchez}, A.~G., {Colavincenzo}, M., {et~al.} 2019, \mnras,
  482, 1786

\bibitem[{{Liu} \& {Haiman}(2016)}]{2016PhRvD..94d3533L}
{Liu}, J., \& {Haiman}, Z. 2016, \prd, 94, 043533

\bibitem[{{Liu} {et~al.}(2015{\natexlab{a}}){Liu}, {Petri}, {Haiman}, {Hui},
  {Kratochvil}, \& {May}}]{2015PhRvD..91f3507L}
{Liu}, J., {Petri}, A., {Haiman}, Z., {et~al.} 2015{\natexlab{a}}, \prd, 91,
  063507

\bibitem[{{Liu} {et~al.}(2015{\natexlab{b}}){Liu}, {Pan}, {Li}, {Shan}, {Wang},
  {Fu}, {Fan}, {Kneib}, {Leauthaud}, {Van Waerbeke}, {Makler}, {Moraes},
  {Erben}, \& {Charbonnier}}]{2015MNRAS.450.2888L}
{Liu}, X., {Pan}, C., {Li}, R., {et~al.} 2015{\natexlab{b}}, \mnras, 450, 2888

\bibitem[{{Liu} {et~al.}(2016){Liu}, {Li}, {Zhao}, {Chiu}, {Fang}, {Pan},
  {Wang}, {Du}, {Yuan}, {Fu}, \& {Fan}}]{2016PhRvL.117e1101L}
{Liu}, X., {Li}, B., {Zhao}, G.-B., {et~al.} 2016, \prl, 117, 051101

\bibitem[{{LSST Science Collaboration} {et~al.}(2009){LSST Science
  Collaboration}, {Abell}, {Allison}, {Anderson}, {Andrew}, {Angel}, {Armus},
  {Arnett}, {Asztalos}, {Axelrod}, {Bailey}, {Ballantyne}, {Bankert},
  {Barkhouse}, {Barr}, {Barrientos}, {Barth}, {Bartlett}, {Becker}, {Becla},
  {Beers}, {Bernstein}, {Biswas}, {Blanton}, {Bloom}, {Bochanski}, {Boeshaar},
  {Borne}, {Bradac}, {Brandt}, {Bridge}, {Brown}, {Brunner}, {Bullock},
  {Burgasser}, {Burge}, {Burke}, {Cargile}, {Chand rasekharan}, {Chartas},
  {Chesley}, {Chu}, {Cinabro}, {Claire}, {Claver}, {Clowe}, {Connolly}, {Cook},
  {Cooke}, {Cooray}, {Covey}, {Culliton}, {de Jong}, {de Vries}, {Debattista},
  {Delgado}, {Dell'Antonio}, {Dhital}, {Di Stefano}, {Dickinson}, {Dilday},
  {Djorgovski}, {Dobler}, {Donalek}, {Dubois-Felsmann}, {Durech},
  {Eliasdottir}, {Eracleous}, {Eyer}, {Falco}, {Fan}, {Fassnacht}, {Ferguson},
  {Fernandez}, {Fields}, {Finkbeiner}, {Figueroa}, {Fox}, {Francke}, {Frank},
  {Frieman}, {Fromenteau}, {Furqan}, {Galaz}, {Gal-Yam}, {Garnavich},
  {Gawiser}, {Geary}, {Gee}, {Gibson}, {Gilmore}, {Grace}, {Green}, {Gressler},
  {Grillmair}, {Habib}, {Haggerty}, {Hamuy}, {Harris}, {Hawley}, {Heavens},
  {Hebb}, {Henry}, {Hileman}, {Hilton}, {Hoadley}, {Holberg}, {Holman},
  {Howell}, {Infante}, {Ivezic}, {Jacoby}, {Jain}, {R}, {Jedicke}, {Jee},
  {Garrett Jernigan}, {Jha}, {Johnston}, {Jones}, {Juric}, {Kaasalainen},
  {Styliani}, {Kafka}, {Kahn}, {Kaib}, {Kalirai}, {Kantor}, {Kasliwal},
  {Keeton}, {Kessler}, {Knezevic}, {Kowalski}, {Krabbendam}, {Krughoff},
  {Kulkarni}, {Kuhlman}, {Lacy}, {Lepine}, {Liang}, {Lien}, {Lira}, {Long},
  {Lorenz}, {Lotz}, {Lupton}, {Lutz}, {Macri}, {Mahabal}, {Mandelbaum},
  {Marshall}, {May}, {McGehee}, {Meadows}, {Meert}, {Milani}, {Miller},
  {Miller}, {Mills}, {Minniti}, {Monet}, {Mukadam}, {Nakar}, {Neill}, {Newman},
  {Nikolaev}, {Nordby}, {O'Connor}, {Oguri}, {Oliver}, {Olivier}, {Olsen},
  {Olsen}, {Olszewski}, {Oluseyi}, {Padilla}, {Parker}, {Pepper}, {Peterson},
  {Petry}, {Pinto}, {Pizagno}, {Popescu}, {Prsa}, {Radcka}, {Raddick},
  {Rasmussen}, {Rau}, {Rho}, {Rhoads}, {Richards}, {Ridgway}, {Robertson},
  {Roskar}, {Saha}, {Sarajedini}, {Scannapieco}, {Schalk}, {Schindler},
  {Schmidt}, {Schmidt}, {Schneider}, {Schumacher}, {Scranton}, {Sebag},
  {Seppala}, {Shemmer}, {Simon}, {Sivertz}, {Smith}, {Allyn Smith}, {Smith},
  {Spitz}, {Stanford}, {Stassun}, {Strader}, {Strauss}, {Stubbs}, {Sweeney},
  {Szalay}, {Szkody}, {Takada}, {Thorman}, {Trilling}, {Trimble}, {Tyson}, {Van
  Berg}, {Vand en Berk}, {VanderPlas}, {Verde}, {Vrsnak}, {Walkowicz}, {Wand
  elt}, {Wang}, {Wang}, {Warner}, {Wechsler}, {West}, {Wiecha}, {Williams},
  {Willman}, {Wittman}, {Wolff}, {Wood-Vasey}, {Wozniak}, {Young}, {Zentner},
  \& {Zhan}}]{2009arXiv0912.0201L}
{LSST Science Collaboration}, {Abell}, P.~A., {Allison}, J., {et~al.} 2009,
  arXiv e-prints, arXiv:0912.0201

\bibitem[{{Martinet} {et~al.}(2018){Martinet}, {Schneider}, {Hildebrandt},
  {Shan}, {Asgari}, {Dietrich}, {Harnois-D{\'e}raps}, {Erben}, {Grado},
  {Heymans}, {Hoekstra}, {Klaes}, {Kuijken}, {Merten}, \&
  {Nakajima}}]{Martinet2018KiDS}
{Martinet}, N., {Schneider}, P., {Hildebrandt}, H., {et~al.} 2018, \mnras, 474,
  712

\bibitem[{Mellier(1999)}]{mellier1999probing}
Mellier, Y. 1999, Annual Review of Astronomy and Astrophysics, 37, 127

\bibitem[{{Munshi} {et~al.}(2019){Munshi}, {Namikawa}, {Kitching}, {McEwen},
  {Takahashi}, {Bouchet}, {Taruya}, \& {Bose}}]{2019arXiv191004627M}
{Munshi}, D., {Namikawa}, T., {Kitching}, T.~D., {et~al.} 2019, arXiv e-prints,
  arXiv:1910.04627

\bibitem[{{Munshi} {et~al.}(2008){Munshi}, {Valageas}, {van Waerbeke}, \&
  {Heavens}}]{2008PhR...462...67M}
{Munshi}, D., {Valageas}, P., {van Waerbeke}, L., \& {Heavens}, A. 2008,
  \physrep, 462, 67

\bibitem[{{Mustafa} {et~al.}(2019){Mustafa}, {Bard}, {Bhimji}, {Luki{\'c}},
  {Al-Rfou}, \& {Kratochvil}}]{2019ComAC...6....1M}
{Mustafa}, M., {Bard}, D., {Bhimji}, W., {et~al.} 2019, Computational
  Astrophysics and Cosmology, 6, 1

\bibitem[{Peel {et~al.}(2016)Peel, Lin, Lanusse, Leonard, Starck, \&
  Kilbinger}]{Peel2016Cosmological}
Peel, A., Lin, C.~A., Lanusse, F., {et~al.} 2016

\bibitem[{{Percival} {et~al.}(2014){Percival}, {Ross}, {S{\'a}nchez},
  {Samushia}, {Burden}, {Crittenden}, {Cuesta}, {Magana}, {Manera}, {Beutler},
  {Chuang}, {Eisenstein}, {Ho}, {McBride}, {Montesano}, {Padmanabhan}, {Reid},
  {Saito}, {Schneider}, {Seo}, {Tojeiro}, \& {Weaver}}]{2014MNRAS.439.2531P}
{Percival}, W.~J., {Ross}, A.~J., {S{\'a}nchez}, A.~G., {et~al.} 2014, \mnras,
  439, 2531

\bibitem[{{Petri} {et~al.}(2015){Petri}, {Liu}, {Haiman}, {May}, {Hui}, \&
  {Kratochvil}}]{2015PhRvD..91j3511P}
{Petri}, A., {Liu}, J., {Haiman}, Z., {et~al.} 2015, \prd, 91, 103511

\bibitem[{{Pires} {et~al.}(2012){Pires}, {Leonard}, \&
  {Starck}}]{2012MNRAS.423..983P}
{Pires}, S., {Leonard}, A., \& {Starck}, J.-L. 2012, \mnras, 423, 983

\bibitem[{{Pires} {et~al.}(2009){Pires}, {Starck}, {Amara},
  {R{\'e}fr{\'e}gier}, \& {Teyssier}}]{2009A&A...505..969P}
{Pires}, S., {Starck}, J.~L., {Amara}, A., {R{\'e}fr{\'e}gier}, A., \&
  {Teyssier}, R. 2009, \aap, 505, 969

\bibitem[{Ribli {et~al.}(2018)Ribli, Pataki, \& Csabai}]{Ribli2018Learning}
Ribli, D., Pataki, B.~Ã., \& Csabai, I. 2018

\bibitem[{{Schneider} {et~al.}(2017){Schneider}, {Ng}, {Dawson}, {Marshall},
  {Meyers}, \& {Bard}}]{2017ApJ...839...25S}
{Schneider}, M.~D., {Ng}, K.~Y., {Dawson}, W.~A., {et~al.} 2017, \apj, 839, 25

\bibitem[{Scoccimarro {et~al.}(1999)Scoccimarro, Zaldarriaga, \&
  Hui}]{Scoccimarro1999Power}
Scoccimarro, R., Zaldarriaga, M., \& Hui, L. 1999, Astrophysical Journal, 527,
  1

\bibitem[{Semboloni {et~al.}(2010)Semboloni, Waerbeke, Heymans, Hamana,
  Colombi, White, \& Mellier}]{Semboloni2010Cosmic}
Semboloni, E., Waerbeke, L.~V., Heymans, C., {et~al.} 2010, Monthly Notices of
  the Royal Astronomical Society Letters, 375, L6

\bibitem[{{Semboloni} {et~al.}(2006){Semboloni}, {Mellier}, {van Waerbeke},
  {Hoekstra}, {Tereno}, {Benabed}, {Gwyn}, {Fu}, {Hudson}, {Maoli}, \&
  {Parker}}]{2006A&A...452...51S}
{Semboloni}, E., {Mellier}, Y., {van Waerbeke}, L., {et~al.} 2006, \aap, 452,
  51

\bibitem[{{Seo} {et~al.}(2012){Seo}, {Sato}, {Takada}, \&
  {Dodelson}}]{2012ApJ...748...57S}
{Seo}, H.-J., {Sato}, M., {Takada}, M., \& {Dodelson}, S. 2012, \apj, 748, 57

\bibitem[{Shan {et~al.}(2017)Shan, Liu, Hildebrandt, Pan, \&
  Qiao}]{Shan2017KiDS}
Shan, H.~Y., Liu, X., Hildebrandt, H., Pan, C., \& Qiao, W. 2017, Monthly
  Notices of the Royal Astronomical Society, 474

\bibitem[{{Shirasaki}(2017)}]{2017MNRAS.465.1974S}
{Shirasaki}, M. 2017, \mnras, 465, 1974

\bibitem[{{Springel}(2005)}]{2005MNRAS.364.1105S}
{Springel}, V. 2005, \mnras, 364, 1105

\bibitem[{Takada \& Jain(2010)}]{Takada2010The}
Takada, M., \& Jain, B. 2010, Monthly Notices of the Royal Astronomical
  Society, 395, 2065

\bibitem[{{Taylor} {et~al.}(2013){Taylor}, {Joachimi}, \&
  {Kitching}}]{2013MNRAS.432.1928T}
{Taylor}, A., {Joachimi}, B., \& {Kitching}, T. 2013, \mnras, 432, 1928

\bibitem[{{Troxel} {et~al.}(2018){Troxel}, {MacCrann}, {Zuntz}, {Eifler},
  {Krause}, {Dodelson}, {Gruen}, {Blazek}, {Friedrich}, {Samuroff}, {Prat},
  {Secco}, {Davis}, {Fert{\'e}}, {DeRose}, {Alarcon}, {Amara}, {Baxter},
  {Becker}, {Bernstein}, {Bridle}, {Cawthon}, {Chang}, {Choi}, {De Vicente},
  {Drlica-Wagner}, {Elvin-Poole}, {Frieman}, {Gatti}, {Hartley}, {Honscheid},
  {Hoyle}, {Huff}, {Huterer}, {Jain}, {Jarvis}, {Kacprzak}, {Kirk}, {Kokron},
  {Krawiec}, {Lahav}, {Liddle}, {Peacock}, {Rau}, {Refregier}, {Rollins},
  {Rozo}, {Rykoff}, {S{\'a}nchez}, {Sevilla-Noarbe}, {Sheldon}, {Stebbins},
  {Varga}, {Vielzeuf}, {Wang}, {Wechsler}, {Yanny}, {Abbott}, {Abdalla},
  {Allam}, {Annis}, {Bechtol}, {Benoit-L{\'e}vy}, {Bertin}, {Brooks},
  {Buckley-Geer}, {Burke}, {Carnero Rosell}, {Carrasco Kind}, {Carretero},
  {Castander}, {Crocce}, {Cunha}, {D'Andrea}, {da Costa}, {DePoy}, {Desai},
  {Diehl}, {Dietrich}, {Doel}, {Fernandez}, {Flaugher}, {Fosalba},
  {Garc{\'\i}a-Bellido}, {Gaztanaga}, {Gerdes}, {Giannantonio}, {Goldstein},
  {Gruendl}, {Gschwend}, {Gutierrez}, {James}, {Jeltema}, {Johnson}, {Johnson},
  {Kent}, {Kuehn}, {Kuhlmann}, {Kuropatkin}, {Li}, {Lima}, {Lin}, {Maia},
  {March}, {Marshall}, {Martini}, {Melchior}, {Menanteau}, {Miquel}, {Mohr},
  {Neilsen}, {Nichol}, {Nord}, {Petravick}, {Plazas}, {Romer}, {Roodman},
  {Sako}, {Sanchez}, {Scarpine}, {Schindler}, {Schubnell}, {Smith}, {Smith},
  {Soares-Santos}, {Sobreira}, {Suchyta}, {Swanson}, {Tarle}, {Thomas},
  {Tucker}, {Vikram}, {Walker}, {Weller}, {Zhang}, \& {DES
  Collaboration}}]{2018PhRvD..98d3528T}
{Troxel}, M.~A., {MacCrann}, N., {Zuntz}, J., {et~al.} 2018, \prd, 98, 043528

\bibitem[{{Van Waerbeke} {et~al.}(2013){Van Waerbeke}, {Benjamin}, {Erben},
  {Heymans}, {Hildebrandt}, {Hoekstra}, {Kitching}, {Mellier}, {Miller},
  {Coupon}, {Harnois-D{\'e}raps}, {Fu}, {Hudson}, {Kilbinger}, {Kuijken},
  {Rowe}, {Schrabback}, {Semboloni}, {Vafaei}, {van Uitert}, \& {Veland
  er}}]{2013MNRAS.433.3373V}
{Van Waerbeke}, L., {Benjamin}, J., {Erben}, T., {et~al.} 2013, \mnras, 433,
  3373

\bibitem[{{Wei} {et~al.}(2018){Wei}, {Li}, {Kang}, {Liu}, {Fan}, {Yuan}, \&
  {Pan}}]{2018MNRAS.478.2987W}
{Wei}, C., {Li}, G., {Kang}, X., {et~al.} 2018, \mnras, 478, 2987

\bibitem[{{Yang} {et~al.}(2011){Yang}, {Kratochvil}, {Wang}, {Lim}, {Haiman},
  \& {May}}]{2011PhRvD..84d3529Y}
{Yang}, X., {Kratochvil}, J.~M., {Wang}, S., {et~al.} 2011, \prd, 84, 043529

\bibitem[{{Yu} {et~al.}(2012){Yu}, {Harnois-D{\'e}raps}, {Zhang}, \&
  {Pen}}]{2012MNRAS.421..832Y}
{Yu}, H.-R., {Harnois-D{\'e}raps}, J., {Zhang}, T.-J., \& {Pen}, U.-L. 2012,
  \mnras, 421, 832

\bibitem[{{Yu} {et~al.}(2016){Yu}, {Zhang}, \& {Jing}}]{2016PhRvD..94h3520Y}
{Yu}, Y., {Zhang}, P., \& {Jing}, Y. 2016, \prd, 94, 083520

\bibitem[{{Yu} {et~al.}(2011){Yu}, {Zhang}, {Lin}, {Cui}, \&
  {Fry}}]{2011PhRvD..84b3523Y}
{Yu}, Y., {Zhang}, P., {Lin}, W., {Cui}, W., \& {Fry}, J.~N. 2011, \prd, 84,
  023523

\bibitem[{{Yuan} {et~al.}(2018){Yuan}, {Liu}, {Pan}, {Wang}, \&
  {Fan}}]{2018ApJ...857..112Y}
{Yuan}, S., {Liu}, X., {Pan}, C., {Wang}, Q., \& {Fan}, Z. 2018, \apj, 857, 112

\bibitem[{{Zhang} {et~al.}(2003){Zhang}, {Pen}, {Zhang}, \&
  {Dubinski}}]{2003ApJ...598..818Z}
{Zhang}, T.-J., {Pen}, U.-L., {Zhang}, P., \& {Dubinski}, J. 2003, \apj, 598,
  818

\end{thebibliography}

\newpage

\end{document}